\documentclass[aps,prd,twocolumn,superscriptaddress,showpacs,preprintnumbers,amsmath,amssymb]{revtex4}

\usepackage{graphicx}% Include figure files
\usepackage{dcolumn}% Align table columns on decimal point
\usepackage{bm}% bold math

\newcommand{\bbar}{\overline{B}{}^{\,0}}
\newcommand{\bdsd}{\overline{B}{}^{\,0}\to D_s^-D^+}
\newcommand{\bdsds}{\overline{B}{}^{\,0}\to D_s^+D_s^-}

\newcommand{\ra}{\!\rightarrow\!}

\begin{document}

\vspace*{-3\baselineskip}
\resizebox{!}{3cm}{\includegraphics{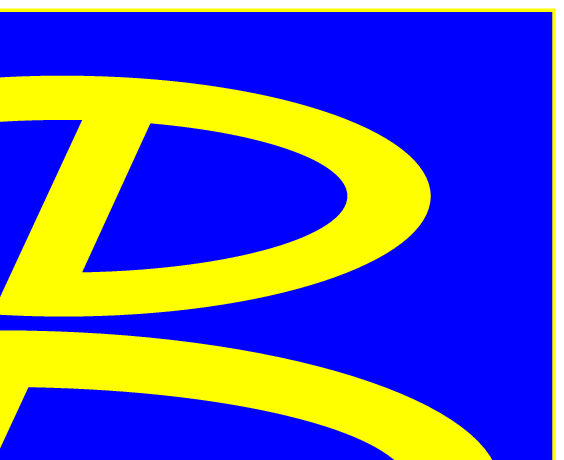}}

\preprint{\vbox{ \hbox{KEK-PREPRINT 2006-80}
                 \hbox{BELLE-PREPRINT 2007-15}}}

\title{\boldmath Improved measurement of $\bbar\to D_s^-D^+$ and 
search for $\bdsds$ at Belle\unboldmath}

\date{\today}

\begin{abstract}
We reconstruct $\bbar\to D_s^-D^+$ decays using a sample of
$449\times 10^6$~$B\overline{B}$ pairs recorded by the Belle 
experiment, and measure the branching fraction to be 
${\cal B}(\bbar\to D_s^-D^+)=\bigl[7.5\pm 0.2({\rm stat})
\pm 0.8({\rm syst})\pm 0.8({\rm {\cal B}'s})\bigr]\times 10^{-3}$.
A search for the related decay $\bdsds$ is also performed.
Since we observe no statistically significant signal an upper 
limit on the branching fraction is set at $3.6\times 10^{-5}$ 
(90\% C.L.).
\end{abstract}

\pacs{13.25.Hw, 14.40.Nd}

\affiliation{Budker Institute of Nuclear Physics, Novosibirsk}
\affiliation{Chiba University, Chiba}
\affiliation{University of Cincinnati, Cincinnati, Ohio 45221}
\affiliation{Department of Physics, Fu Jen Catholic University, Taipei}
%%%\affiliation{Justus-Liebig-Universit\"at Gie\ss{}en, Gie\ss{}en}
\affiliation{The Graduate University for Advanced Studies, Hayama}
\affiliation{Gyeongsang National University, Chinju}
\affiliation{Hanyang University, Seoul}
\affiliation{University of Hawaii, Honolulu, Hawaii 96822}
\affiliation{High Energy Accelerator Research Organization (KEK), Tsukuba}
%%%\affiliation{Hiroshima Institute of Technology, Hiroshima}
\affiliation{University of Illinois at Urbana-Champaign, Urbana, Illinois 61801}
\affiliation{Institute of High Energy Physics, Chinese Academy of Sciences, Beijing}
\affiliation{Institute of High Energy Physics, Vienna}
\affiliation{Institute of High Energy Physics, Protvino}
\affiliation{Institute for Theoretical and Experimental Physics, Moscow}
\affiliation{J. Stefan Institute, Ljubljana}
\affiliation{Kanagawa University, Yokohama}
\affiliation{Korea University, Seoul}
%%%\affiliation{Kyoto University, Kyoto}
\affiliation{Kyungpook National University, Taegu}
\affiliation{Swiss Federal Institute of Technology of Lausanne, EPFL, Lausanne}
\affiliation{University of Ljubljana, Ljubljana}
\affiliation{University of Maribor, Maribor}
\affiliation{University of Melbourne, School of Physics, Victoria 3010}
\affiliation{Nagoya University, Nagoya}
\affiliation{Nara Women's University, Nara}
\affiliation{National Central University, Chung-li}
\affiliation{National United University, Miao Li}
\affiliation{Department of Physics, National Taiwan University, Taipei}
\affiliation{H. Niewodniczanski Institute of Nuclear Physics, Krakow}
\affiliation{Nippon Dental University, Niigata}
\affiliation{Niigata University, Niigata}
\affiliation{University of Nova Gorica, Nova Gorica}
\affiliation{Osaka City University, Osaka}
\affiliation{Osaka University, Osaka}
\affiliation{Panjab University, Chandigarh}
\affiliation{Peking University, Beijing}
%%%\affiliation{University of Pittsburgh, Pittsburgh, Pennsylvania 15260}
%%%\affiliation{Princeton University, Princeton, New Jersey 08544}
\affiliation{RIKEN BNL Research Center, Upton, New York 11973}
\affiliation{Saga University, Saga}
\affiliation{University of Science and Technology of China, Hefei}
\affiliation{Seoul National University, Seoul}
\affiliation{Shinshu University, Nagano}
\affiliation{Sungkyunkwan University, Suwon}
\affiliation{University of Sydney, Sydney, New South Wales}
\affiliation{Tata Institute of Fundamental Research, Mumbai}
\affiliation{Toho University, Funabashi}
\affiliation{Tohoku Gakuin University, Tagajo}
\affiliation{Tohoku University, Sendai}
\affiliation{Department of Physics, University of Tokyo, Tokyo}
\affiliation{Tokyo Institute of Technology, Tokyo}
\affiliation{Tokyo Metropolitan University, Tokyo}
\affiliation{Tokyo University of Agriculture and Technology, Tokyo}
%%%\affiliation{Toyama National College of Maritime Technology, Toyama}
%%%\affiliation{University of Tsukuba, Tsukuba}
\affiliation{Virginia Polytechnic Institute and State University, Blacksburg, Virginia 24061}
\affiliation{Yonsei University, Seoul}
  \author{A.~Zupanc}\affiliation{J. Stefan Institute, Ljubljana} % Ljubljana
  \author{K.~Abe}\affiliation{High Energy Accelerator Research Organization (KEK), Tsukuba} % KEK
  \author{K.~Abe}\affiliation{Tohoku Gakuin University, Tagajo} % TohokuGakuin
% \author{N.~Abe}\affiliation{Tokyo Institute of Technology, Tokyo} % TIT
% \author{I.~Adachi}\affiliation{High Energy Accelerator Research Organization (KEK), Tsukuba} % KEK
  \author{H.~Aihara}\affiliation{Department of Physics, University of Tokyo, Tokyo} % Tokyo
  \author{D.~Anipko}\affiliation{Budker Institute of Nuclear Physics, Novosibirsk} % BINP
% \author{K.~Aoki}\affiliation{Nagoya University, Nagoya} % Nagoya
  \author{K.~Arinstein}\affiliation{Budker Institute of Nuclear Physics, Novosibirsk} % BINP
% \author{Y.~Asano}\affiliation{University of Tsukuba, Tsukuba} % Tsukuba
% \author{T.~Aso}\affiliation{Toyama National College of Maritime Technology, Toyama} % Toyama
  \author{V.~Aulchenko}\affiliation{Budker Institute of Nuclear Physics, Novosibirsk} % BINP
  \author{T.~Aushev}\affiliation{Swiss Federal Institute of Technology of Lausanne, EPFL, Lausanne}\affiliation{Institute for Theoretical and Experimental Physics, Moscow} % ITEP
% \author{T.~Aziz}\affiliation{Tata Institute of Fundamental Research, Mumbai} % Tata
  \author{S.~Bahinipati}\affiliation{University of Cincinnati, Cincinnati, Ohio 45221} % Cincinnati
  \author{A.~M.~Bakich}\affiliation{University of Sydney, Sydney, New South Wales} % Sydney
% \author{V.~Balagura}\affiliation{Institute for Theoretical and Experimental Physics, Moscow} % ITEP
% \author{Y.~Ban}\affiliation{Peking University, Beijing} % Peking
% \author{S.~Banerjee}\affiliation{Tata Institute of Fundamental Research, Mumbai} % Tata
% ...
% BINP
% \author{T.~Ziegler}\affiliation{Princeton University, Princeton, New Jersey 08544} % Princeton
% \author{D.~Z\"urcher}\affiliation{Swiss Federal Institute of Technology of Lausanne, EPFL, Lausanne} % Lausanne
\collaboration{The Belle Collaboration}

\maketitle

\tighten

\section{Introduction}
Several decay modes of $B$ mesons with a $D^+_s$ in the
final state have been measured at the $B$-factories.
The amplitudes governing these decays are interesting
because none of the constituent flavors of the $D^+_s$
are present in the initial state. For example, the decays
$\bbar\ra D^+_s K^-$~\cite{Krokovny:2002pe, Aubert:2002vg} 
and $\bbar\ra D^{\ast}_{s0}(2317)^+ K^-$~\cite{Drutskoy:2005zr}, 
observed with branching fractions in the range $10^{-5}-10^{-4}$, can 
proceed via a $b\overline{d}\ra c\overline{u}$ $W$-exchange diagram. 
Here we study the related decays $\bbar\ra D^+_s D^-_s$ and 
$\bbar\ra D^-_s D^+$. The former proceeds via Cabibbo-suppressed 
$W$-exchange and has not yet been observed; theoretical calculations 
predict a branching fraction ranging from 
$\sim$\,$8\times 10^{-5}$~\cite{Li:2003az} up to 
$\sim$\,$3\times 10^{-4}~$~\cite{Eeg:2005au}.
The latter of the two above decays proceeds via a Cabibbo-favored tree 
diagram; the ratio of its branching fraction to that for $\bbar\ra D^+\pi^-$
can be used to test the factorization hypothesis for exclusive non-leptonic 
decays of $B$ mesons~\cite{Kim:2001cj}. However, previous measurements of 
${\cal B}(\bbar\ra D^+_s D^-)$~\cite{Bortoletto:1991kz, Albrecht:1991pa, Gibaut:1995tu, Aubert:2006nm} 
have large uncertainties, which limit the usefulness of this method at present.

In this paper we report an improved measurement of $\bbar\to D_s^-D^+$ decays and a search for $\bbar\to
D_s^+D_s^-$ decays with the Belle detector~\cite{Belle_det} at the
KEKB asymmetric-energy $e^+e^-$ collider~\cite{KEKB_acc}. 
Charge conjugate modes are implied throughout this paper. 
The results are based on a $414$ fb$^{-1}$ data sample collected at the center-of-mass (CM)
energy of the $\Upsilon(4S)$ resonance, corresponding to 
$(449.3 \pm 5.7)\times 10^{6}~B\overline{B}$ pairs. We assume equal production of $B^0\bbar$ and $B^+B^-$ pairs.
To study backgrounds, we use a Monte Carlo (MC)
simulated sample~\cite{qq98} of $\Upsilon(4S)\to B\overline{B}$ 
events and continuum events, $e^+e^-\to q\overline{q}$ ($q=u$, $d$, $s$ and $c$ quarks).

The Belle detector is a large-solid-angle magnetic spectrometer that consists of a multi-layer silicon
vertex detector (SVD), a 50-layer central drift chamber (CDC), an array of aerogel threshold
Cherenkov counters (ACC), a barrel-like arrangement of time-of-flight scintillation counters
(TOF), and an electromagnetic calorimeter (ECL) comprised of CsI(Tl) crystals located
inside a superconducting solenoid coil that provides a $1.5$ T magnetic field. An iron flux-return
located outside of the coil is instrumented to detect $K^0_L$ mesons and to identify muons (KLM).
The detector is described in detail in Ref.~\cite{Belle_det}. Two different inner detector
configurations were used. For the first 152 million $B\overline{B}$ pairs, a $2.0$ cm radius
beampipe and a 3-layer silicon vertex detector were used; for the latter 297 million $B\overline{B}$ pairs,
a $1.5$ cm radius beampipe, a 4-layer silicon detector and a small-cell inner drift chamber were
used~\cite{Natkaniec:2006rv}.

\section{Reconstruction}

Charged tracks are selected with
loose requirements on their impact parameters relative to the interaction
point (IP) and the transverse momentum of the tracks.
For charged particle identification (PID) we combine 
information from the CDC, TOF and ACC counters into a likelihood ratio 
${\cal{L}}(K)/({\cal{L}}(K)+{\cal{L}}(\pi))$~\cite{Nakano:2002jw}. 
A selection imposed on this ratio results in a typical kaon (pion) identification efficiency ranging from 92\% to
97\% (94\% to 98\%) for various decay modes, while 2\% to 15\% (4\% to 8\%) of kaon (pion) candidates are 
misidentified pions (kaons).

We use the $D_s^-\rightarrow\phi\pi^-$, $K^{*0}K^-$ and $K^0_SK^-$ modes to reconstruct $D_s^-$ mesons and
$D^+\rightarrow K^+K^-\pi^+$, $K^- \pi^+\pi^+$, and $K^0_S\pi^+$ for the $D^+$ mesons, where the $\phi$,
$K^{*0}$ and $K^0_S$ decay to $K^+K^-$, $K^+\pi^-$ and $\pi^-\pi^+$,
respectively.
Combinations of oppositely-charged kaons with 
$|m_\phi-M_{K^+K^-}|<20$~MeV/$c^2$ and of oppositely-charged kaons and
pions with $|m_{K^{\ast 0}}-M_{K^+\pi^-}|<85$~MeV/$c^2$, originating
from a common vertex, are retained as $\phi$ and $K^{\ast 0}$
candidates, where $m_\phi$ and $m_{K^{\ast 0}}$ are the nominal masses of
the two mesons~\cite{Yao:2006px}.
Neutral kaons ($K^0_S$) are reconstructed using 
pairs of oppositely-charged tracks 
that have an invariant mass within 30 MeV$/c^2$ of the nominal $K^0$ mass, 
and originate from a common vertex, displaced from the IP.
All $D_{(s)}$ candidates with invariant masses within a $4~\sigma$
($4.5~\sigma$) interval around the  nominal $D_s$ ($D$) mass are considered for further
analysis, where $D_s$ ($D$) signal resolutions ($\sigma$) range from $3.6$~MeV/$c^2$ to $4.2$~MeV/$c^2$ ($3.7$~MeV/$c^2$ to
$4.1$~MeV/$c^2$). 
A decay vertex fit with a mass constraint is applied to the selected $D_{(s)}$ candidates to improve their momentum resolution.
For the decay $\bdsds$ we also add an additional
constraint on the value of the cosine of a helicity angle, 
$|\cos\theta_h| > 0.05~(0.25)$ for the $D_s^-\to\phi\pi^- ~(K^{*0}K^-)$ decay mode,
where $\theta_h$ is defined as the angle between the
direction of the $D_s^{-}$ and the $K^+$ originating from the vector-meson ($\phi$ or 
$K^{\ast 0}$) in the vector-meson rest frame. 
The distribution in $\cos\theta_h$ is expected to be proportional to $\cos^2\theta_h$ for the signal and uniform for 
the combinatorial background.

Pairs of $D_s^-$ and $D^+_{(s)}$ meson candidates are combined to form $\bbar$ meson candidates. These
are identified by their CM energy difference,
$\Delta E=E_B^{\rm CM}-E_{\rm beam}^{\rm CM}$,
and the beam-energy constrained mass,
$M_{\rm bc}=\sqrt{(E_{\rm beam}^{\rm CM})^2-(p_B^{\rm CM})^2}$, 
where $E_{\rm beam}^{\rm CM} = \sqrt{s}/2$ is the CM beam energy and
$E_B^{\rm CM}$ and $p_B^{\rm CM}$ are
the reconstructed energy and momentum of the $B$ meson candidate in
the CM frame. The signal region is $5.272$ GeV$/c^2$ $\leq$ $M_{\rm bc}$ $\leq$ $5.285$ GeV$/c^2$ 
for the $\bdsd$, and 
$5.274$ GeV$/c^2$ $\leq$ $M_{\rm bc}$ $\leq$ $5.284$ GeV$/c^2$ and
$|\Delta E|\leq$ $0.013$ GeV for the $\bdsds$ decays. 

To suppress the large combinatorial background dominated by the two-jet-like $e^+e^-\to q\overline{q}$
continuum process, variables characterizing the event topology are used. We
require the ratio of the second to zeroth Fox-Wolfram
moments~\cite{Fox:1978vu}, $R_2<0.3$ and 
the thrust value of the event, $T<0.8$.
Simulation shows that this selection retains more than 95\% of
$B\overline{B}$ events and 
rejects about 55\% of
$c\overline{c}$ events and 65\% of $u\overline{u}$, $d\overline{d}$ 
and $s\overline{s}$ events. 

The above selection criteria and signal regions are determined by maximizing the figure of merit (FoM), 
$S/\sqrt{S+B}$, where $S$ and $B$ are the numbers of 
signal and background events determined from MC. For optimization of the FoM we assume 
${\cal B}(\bdsds)=2\times 10^{-4}$.

The fraction of events with more than one $\bdsd$ ($\bdsds$) candidate is 4.9\% (2.8\%).
As the best candidate we select the one with the minimal $\chi^2 = \chi^2(D_s^-) +
\chi^2(D_{(s)}^+)$ value, where $\chi^2(D_s^-)$ and $\chi^2(D_{(s)}^+)$ are $\chi^2$'s 
of the mass-constrained vertex fit.

\boldmath
\section{ $\bbar\to D_s^-D^+$ decays}
\unboldmath
\begin{figure}[t]
\begin{center}
\includegraphics[width=0.45\textwidth]{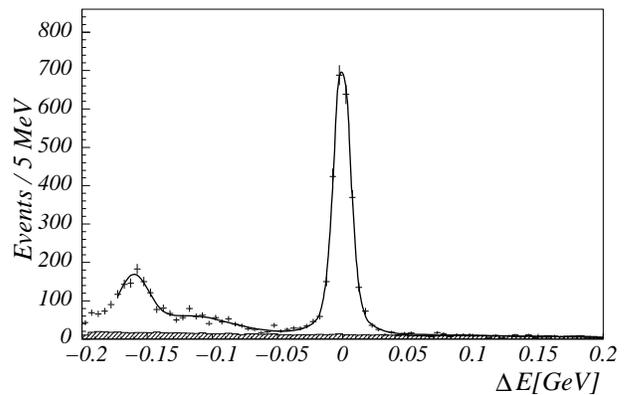}
\end{center}
\caption{$\Delta E$ distribution for reconstructed $\bbar\to D_s^-D^+$ 
events in the $M_{\rm bc}$ signal region. The curve shows the result of 
the fit. The normalized distribution for the events in the sidebands
of both $D_s$ and $D$ invariant masses is shown 
as the hatched histogram.} 
\label{fig_8}
\end{figure}
The $\Delta E$ distribution of events in the $M_{\rm bc}$ signal
region, obtained after 
applying all selection criteria described above is shown in
Fig.~\ref{fig_8}. Apart from the signal peak at $\Delta E =0$,
contributions from two other specific decay modes were identified using the MC: $\overline{B}{}^0\to D_s^{\ast -}D^{+}$ 
and $\overline{B}{}^0\to D_s^{-}D^{\ast +}$. 
These events cluster around $\Delta E = -0.16 {\rm~GeV}$ and $-0.10 {\rm~GeV}$ due to
the unreconstructed $\pi^0$ or $\gamma$ from the $D_{(s)}^\ast$ meson.

The $\Delta E$ distribution is described by two Gaussians with the same mean for the signal, 
two Gaussians for the $\overline{B}{}^0\to D_s^{\ast -}D^{+}$, $D_s^{-}D^{\ast +}$ background events,
and a linear function for the rest of the background.
The normalizations, positions and widths of the Gaussians are free
parameters of the binned likelihood fit.
~The solid line in Fig.~\ref{fig_8} shows the result
of the fit.
The positions and widths of the 
$\overline{B}{}^0\to D_s^{\ast -}D^{+}$, $D_s^{-}D^{\ast +}$ background components agree with the
values expected from the MC. 
In addition, we perform separate fits to the $\Delta E$ distributions for each $D_s$ 
decay mode using the same function with the widths and means of all four 
Gaussian functions fixed to the values obtained by the overall $\Delta E$ fit.

We use events in the $D_s$ and $D$ meson invariant mass sidebands 
in order to check for 
peaking backgrounds. 
For this check the masses of $D_s$ and $D$ candidates are not constrained
to their nominal masses. The $D$($D_{s}$) invariant mass sidebands are $\pm 200$~MeV/$c^2$ intervals around $D$($D_{s}$) nominal mass, excluding the $D$($D_{s}$) signal region. 
Due to common final states used to reconstruct $D$ and $D_s$ candidates we
exclude the $D_{s}$($D$) signal regions and a $\pm 27$~MeV/$c^2$ $D^{\ast +}$ mass region from $D$($D_{s}$) sidebands.
The $\Delta E$ and $M_{\rm bc}$ distributions obtained by
simultaneously using events in the sidebands of 
both the $D$ and $D_s$ mesons are in agreement with the observed
combinatorial background under the $\bbar\to D_s^-D^+$ signal. 
A significant signal is present only in the $D_s$ sideband, 
for $D_s$'s reconstructed in the $D_s^-\to K^{\ast 0}K^-$ decay mode. 
This is due to the three-body 
$\overline{B}{}^0\to D^+ K^{\ast 0} K^-$ decay, reported in Ref.~\cite{Drutskoy:2002ib}. 
The fraction of these events in the signal peak was evaluated by fitting the $\Delta E$ distribution
in the $D_s$ sideband.
We observe no peaking background when using the $D$ mass sideband. 
The signal in Fig.~\ref{fig_8} also includes contributions from
$D_s^+\to K^+K^-\pi^+$, $f_0(980) \pi^+$ and 
$\overline{K}{}^{\ast}_0(1430)^0 K^+$, which all have a common final state, as well as a small contribution ($0.4\%$) from
 $D^+ \to K^- \pi^+ \pi^+$ decays, where one of the $\pi^+$ decays in-flight to a $\mu^+$ and $\nu_{\mu}$ and the $\mu^+$ is
misidentified as the $\pi^+$. We evaluate these fractions using simulated events. The contribution of these decays is around 
five times larger than the contribution of $\overline{B}{}^0\to D^+ K^{\ast 0} K^-$ decays.
We take into account the relative contributions of individual $D_s$ and $D$
decay modes and determine the overall fraction of peaking background events ($r$) to be ($11.3\pm 2.6$)\%.
The uncertainty includes the
statistical uncertainty in $D_s$ sideband fits, non-uniformity of 
$M(K^{\ast 0} K^-)$ in $\overline{B}{}^0\to D^+ K^{\ast 0} K^-$ decays, 
limited MC statistics and uncertainties in
the corresponding branching fractions~\cite{Yao:2006px}. 

The signal yield for
$\bbar\to D_s^-D^+$ is thus $N=(1-r)N_{\rm peak}=2230\pm 56({\rm stat})$, where $N_{\rm peak}$ is the number
of events in the signal peak obtained from the fit to the $\Delta E$ distribution (Fig.~\ref{fig_8}).

\boldmath
\section{$\bdsds$ decays}
\begin{figure}[t]
\begin{center}
\includegraphics[width=0.45\textwidth]{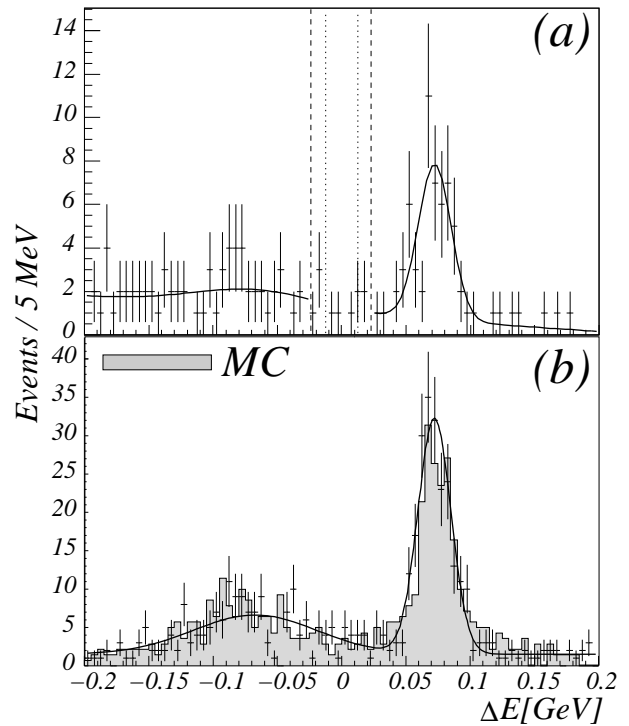}
\end{center}
\caption{(a) $\Delta E$ distribution for the $\bbar\to
D_s^+D_s^-$ decay mode. Two vertical dashed lines show the interval excluded
from the fit, as described in the text, and two dotted lines show the $\Delta E$ signal
region. (b) $\Delta E$ distribution for reconstructed events obtained 
by inverting the kaon identification requirements in data and in the MC sample.}
\label{fig_12}
\end{figure}

\unboldmath
The $\Delta E$ distribution for $\bdsds$ decays obtained after applying 
all selection criteria described above is shown in Fig.~\ref{fig_12}(a). 
The expected width of the narrower signal Gaussian, which describes $82\%$ of the events, is $5.8$~MeV. This value is
obtained from the MC sample and rescaled by a factor obtained after a comparison of parameters from $\bdsd$ data 
and MC samples. The $\Delta E$ signal region includes around $89\%$ of the signal.

While the $\Delta E$ distribution of the combinatorial background is well 
described by a first order polynomial, there is a significant 
cross-feed contribution from $\bdsd$, $D_s^{\ast -}D^{+}$, and $D_s^{ -}D^{\ast +}$  decays, 
where the $D^+$ decays into a $K^-\pi^+\pi^+$ or $\overline{K}{}^0\pi^+$ final state and 
one of the pions is misidentified as a kaon. Figure~\ref{fig_12}(b) shows the $\Delta E$ 
distribution of these cross-feed events, as obtained in both data and MC samples  by 
selecting one of the kaon tracks in the $D_s$ decay chain with a pion PID requirement. Events peaking around 
$0.075$~GeV are due to 
$\overline{B}{}^0\to D_s^{ -}D^{ +}$ decays, while the events clustering around $-0.1$~GeV
are due to $\overline{B}{}^0\to D_s^{\ast -}D^{ +}$ and $\overline{B}{}^0\to D_s^{ -}D^{\ast +}$
decays without a reconstructed $\pi^0$ or a photon.
The $\Delta E$ distribution of 
cross-feed events is described by the sum of two Gaussian functions and a
constant. The solid line in Fig.~\ref{fig_12}(b) shows the result of the fit.
The widths and means of the two Gaussian functions are statistically consistent with the values obtained from MC.

The expected number of background events populating the $\Delta E$ signal region is determined by a binned likelihood 
fit to the $\Delta E$ distribution sidebands ($|\Delta E| > 24$~MeV region indicated by the two vertical dashed lines in Fig.~\ref{fig_12}(a)).
While normalizations are free parameters of the fit, the widths and means of the two Gaussian functions are fixed to the values obtained from
 a fit to the $\Delta E$ distribution of the misidentified data (Fig.~\ref{fig_12}(b)). The fit result is then integrated
across the $\Delta E$ signal region (indicated by the two dotted lines in Fig.~\ref{fig_12}(a)) to obtain the
number of background events, $b = 6.7 \pm 0.8({\rm stat}) \pm 0.5({\rm syst})$, where
the systematic error is evaluated by varying
values of the fixed fit parameters by one standard deviation. Since only three events are observed in the $\Delta E$ signal region, 
the result for $b$ indicates that there is no statistically significant signal present in
this $\Delta E$ interval. Thus the expected $3.5\%$ tail of the signal, which might populate 
the fitted region (parameterized as background only), can be safely neglected.

The average efficiency of the selection criteria $\epsilon (D_sD_s)=\sum_{i,j}\epsilon_{i,j}{\cal B}(D_{si}){\cal
B}(D_{sj})=(2.45\pm 0.46)\times 10^{-4}$ is evaluated from MC, where the intermediate branching fractions ${\cal B}(D_{s}\to\phi\pi)$ and ${\cal B}(D_{s}\to K^{*0} K)$ are taken from Ref.~\cite{Yao:2006px}, and  ${\cal B}(D_{s}\to K^0_S K)$ is taken from Ref.~\cite{Adam:2006me}.

To check for a possible peaking background we use events in the $D_s$ mass sidebands.
No peaking structures are observed in any of the $M_{\rm bc}$-$\Delta E$ distributions.

\section{Results}
\begin{table}[t!]
\caption{Sources of systematic uncertainty in ${\cal B}(\bbar\to D_s^-D^+)$ and 
${\cal B}(\bdsds)$ measurements. }
\label{tab_4}
\renewcommand{\arraystretch}{1.2}
\begin{tabular}{ccc}
\hline \hline
Systematics &  $\bbar\to D_s^-D^+[\% ]$ & $\bdsds[\% ]$\\
\hline
${\cal B}$'s of $D_s$ and $D$ mesons			& $10.1$ 	& $18.6$ \\ 
Tracking					& $6.0$ 	& $6.0$ \\ 
$PID(K^{\pm}/\pi^{\pm})/K^0_S$ $\epsilon$       & $7.4$ 	& $8.3$\\ 
MC statistics 					& $1.7$ 	& $3.9$\\
Signal window					& $1.0$		& $2.1$\\ 
Signal fraction ($1-r$)				& $2.9$  	& \\  
Fitting model 					& $1.9$ 	& included in $b$\\ 
$N(B\overline{B})$ 				        & $1.3$  	& $1.3$\\ 
\hline
Total 						& $14.5$ 	& $21.7$\\
\hline \hline
\end{tabular}
\end{table}
We consider several sources of systematic
uncertainty as listed in Table~\ref{tab_4}. The largest contribution
arises from an imprecise knowledge of the intermediate branching fractions of $D_s$ and $D$ mesons and amounts to $\pm 10.1\%$ 
($\pm 18.6\%$) for $\bdsd$ ($\bdsds$) decay mode~\cite{Yao:2006px, Adam:2006me}.
A $1\%$ relative error for each of the charged tracks used in the
reconstructed final states 
is assigned due to the
uncertainty in tracking efficiency determined using partially reconstructed $D^{\ast}$'s. 
The particle identification efficiency has a relative uncertainty of $1.4\%$ per charged kaon
and $0.8\%$ per charged pion, determined from $D^{\ast +}\to
D^0\pi^+$, $D^0\to K^-\pi^+$ decays. 
The relative error for each reconstructed $K^0_S$ in the final state is $4.5\%$.
A $1.7\%$ ($3.9\%$) uncertainty is due to the limited statistics of the MC
sample used for the efficiency calculation. Since the efficiency is evaluated for the signal region,
we assign an additional $1\%$  ($2.1\%$) uncertainty due to
the small possible difference in the signal resolution between data and MC samples.
A $2.9\%$ uncertainty is due to the imprecise knowledge of the fraction of true signal events, ($1-r$), in the data sample.
Systematic uncertainty arising from the description of the $\Delta E$ distribution is evaluated by comparing the known number of
reconstructed $\bbar\to D_s^-D^+$ events in the simulated sample with the fitted yield and 
is found to be $1.9\%$. Finally, the
uncertainty in the number of $B\overline{B}$ events ($1.3\%$) is taken into account. 
The sum in quadrature of the individual
contributions gives a systematic error of $14.5\%$ for a ${\cal B}(\bbar\to D_s^-D^+)$ and $21.7\%$
for a ${\cal B}(\bdsds)$ measurement, excluding the uncertainty due to the fitting model included in $b$.

\begin{table*}[t!]
\caption{Results on the fitted numbers of events in the signal peak and branching fractions for $\bbar\to D_s^-D^+$ decay mode. 
The peaking background fraction, $r$, is given for each $D_s$ decay mode in the second column. 
The efficiencies include intermediate branching fractions ($\epsilon (D_sD)=\sum_{j}\epsilon_{j}{\cal B}(D_{s}){\cal B}(D_{j})$), 
which are taken for all three $D$ and $D_s\to \phi\pi$ and $K^{*0}K$ decay modes from Ref.~\cite{Yao:2006px}, and that for 
$D_s\to K^0_SK$ is taken from Ref.~\cite{Adam:2006me}.
In the last column the dominant sources of systematic uncertainty, the $D_s$ branching fractions, ${\cal B}_{D_sX}$, are factored out.
Quoted uncertainties for ${\cal B}$ are statistical, systematic and uncertainty due to the imprecise knowledge of intermediate branching fractions, respectively.}
\label{tab_dsd}
\renewcommand{\arraystretch}{1.2}
\begin{tabular}{lr@{~$\pm$~}l@{~~~}r@{~$\pm$~}l@{~~~}c@{~~~}c@{~~~}c}
\hline \hline
Mode										& \multicolumn{2}{c}{$N_{peak}$}	& \multicolumn{2}{c}{$r[\%]$}& $\epsilon$ $[10^{-4}]$		& ${\cal B}$ $[10^{-3}]$			& ${\cal B}\cdot{\cal B}_{D_sX}$ $[10^{-4}]$  \\ \hline
$D_s^-\to \phi\pi^-$, $\phi\to K^+K^-$			& $1112$	& $35$					&$12.9$	& $4.5$				& $2.8 \pm 0.4$					& $7.8 \pm 0.2 \pm 0.9 \pm 1.0$	& $1.68 \pm 0.05 \pm 0.19 \pm 0.06$  \\
$D_s^-\to K^{*0}K^-$, $K^{*0}\to K^+\pi^-$ 	& $961$		& $33$					&$14.5$	& $4.3$				& $2.5 \pm 0.5$					& $7.3 \pm 0.3 \pm 0.8 \pm 1.5$	& $1.83 \pm 0.06 \pm 0.21 \pm 0.07$ \\ 
$D_s^-\to K^0_SK^-$, $K^0_S\to \pi^+\pi^-$	& $441$		& $22$					&$0.4$	& $2.2$				& $1.3 \pm 0.1$					& $7.3 \pm 0.4 \pm 0.9 \pm 0.6$	& $0.76 \pm 0.04 \pm 0.09 \pm 0.03$ \\ \hline
Combined									& $2514$	& $64$					&$11.3$	& $2.6$				& $6.6 \pm 0.7$					& $7.5 \pm 0.2 \pm 0.8 \pm 0.8$	&  \\ 
\hline \hline
\end{tabular}
\end{table*}

The number of signal $\bbar\to D_s^-D^+$ events, $N_{D_sD}$, is converted into a 
branching fraction using the MC efficiency $\epsilon(D_sD)$ and the  
number of $B\overline{B}$ events.
The measured branching fraction is given in Table~\ref{tab_dsd}.

We use the world average of ${\cal B}(\bbar\ra D^+\pi^-)$ \cite{Yao:2006px} and calculate the ratio
\begin{equation}
	R_{D_s/\pi}^{\rm ex.} = \frac{{\cal B}(\bdsd)}{{\cal B}(\bbar\ra D^+\pi^-)} = 2.65 \pm 0.42.
\end{equation}
Before comparing  this result to the numerical prediction of $R_{D_s/\pi}$ given in Ref.~\cite{Kim:2001cj} --- in which the calculation
is performed in the generalized factorization scheme and includes penguin effects --- we rescale it by a factor 
$(f_{D_s}^{\rm new}/f_{D_s}^{\rm old})^2$, where $f_{D_s}^{\rm new}$ is the average value of $D_s$ meson decay constant
given in Refs.~\cite{Yao:2006px, Artuso:2006kz} and $f_{D_s}^{\rm old}$ is the value used in the original calculation. 
The expected value is $R_{D_s/\pi}^{\rm th.} = 3.12\pm 0.35$, where the uncertainty originates
from the dependence on the decay constant $f_{D_s}$ and form-factors, the former being the main source.
The ratio $R_{D_s/\pi}^{\rm ex.}/R_{D_s/\pi}^{\rm th.} = 0.85\pm 0.13(\rm ex.)\pm 0.09(\rm th.)$ is consistent with unity. 
If one does not include the penguin contributions~\cite{Kim:2001cj} to the amplitude for $\bdsd$ decay, the above ratio would be $0.61\pm 0.10(\rm ex.)\pm 0.07(\rm th.)$.

We observe no statistically significant signal in the $\bdsds$ decay mode. 
The central value for the measured branching fraction is 
$[-3.4 \pm 1.6({\rm stat}) \pm 0.6 ({\rm syst}) \pm 0.6({\rm {\cal B}'s})]\times 10^{-5}$.
We infer an upper limit on the ${\cal B}(\bdsds)$ from the total measured number of
reconstructed events and the number of background events in the $\Delta E$ signal region 
($n_0 = 3$ and $b = 6.7 \pm 0.9$, respectively),
and the measured sensitivity, $S_0 = N_{B^0\overline{B}{}^0}\cdot \epsilon(D_sD_s) = (110 \pm 24) \times 10^{3}$. The latter error includes all systematic uncertainties given in Table~\ref{tab_4}. To estimate the upper
limit we use Bayes's theorem with a flat-prior for the signal following the prescription in~(Section 32.3.1 in Ref.~\cite{Yao:2006px}):
\begin{widetext}
\begin{equation}
	p({\cal B}|n_0, b, \sigma_b, S_0,\sigma_{S}) = 
	\frac{\int_{-\infty}^{\infty}\int_{-\infty}^{\infty}L(n_0|{\cal B},\mu_b){\cal G}(S|S_0,\sigma_S)\pi({\cal B},\mu_b|b,\sigma_b)d\mu_b dS}{\int_{-\infty}^{\infty}\int_{-\infty}^{\infty}\int_{-\infty}^{\infty}L(n_0|{\cal B}',\mu_b'){\cal G}(S'|S_0,\sigma_S)\pi({\cal B}',\mu_b'|b,\sigma_b)d{\cal B}'d\mu_b'dS'}.
	\label{Bayes}
\end{equation}
\end{widetext}
The number of observed events $n_0$ is Poisson distributed around the sum of $\mu_s$ and $\mu_b$:
$L(n_0|\mu_s,\mu_b)=1/n_0!\,(\mu_s+\mu_b)^{n_0}e^{-(\mu_s+\mu_b)}$,
where $\mu_s$ and $\mu_b$ are the expected number of signal and
background events, respectively. 
In particular
$\mu_s$ can be written as $\mu_s = {\cal B}\cdot S$,
where ${\cal B}$ and $S$ are true values of ${\cal B}(\bdsds)$ and the sensitivity 
$N_{B^0\overline{B}{}^0}\cdot \epsilon(D_sD_s)$, respectively. 
The true value of $S$ can only take non-negative values and is Gaussian distributed
around $S_0$ with variance $\sigma_S$. Hence 
${\cal{G}}(S|S_0,\sigma_S)$ is a Gaussian function with a cut-off for $S<0$.
The prior probability density $\pi({\cal B},\mu_b|b,\sigma_b)$ 
is assumed to be factorizable, 
$\pi({\cal B},\mu_b|b,\sigma_b)=P({\cal B}){\cal{G}}(\mu_b|b,\sigma_b)$. For $P(\cal{B})$ 
we use 
a flat-prior, and ${\cal{G}}(\mu_b|b,\sigma_b)$ is again a Gaussian function
centered at $b$, with a width of $\sigma_b$ and with a cut-off for $b<0$.

Integrating out the nuisance parameters $S$ and $\mu_b$
we obtain the posterior $p({\cal B}|n_0, b, \sigma_b,
S_0,\sigma_{S})$, which already takes into account the
statistical error on $b$, the systematic error due to the
parameterization of $\Delta E$ distribution in the fit, 
and systematic uncertainties on the efficiency and on the number of
$B\overline{B}$ pairs. 
The 90\% C. L. upper limit on ${\cal B}(\bdsds)$ following
from this posterior is found to be
\[
 {\cal B}(\bdsds) ~\le~3.6\times 10^{-5}~~{\rm at}~90\% ~{\rm C.L.}.
\]

\section{Conclusions}
In conclusion, we have measured the branching fraction for $\bbar\to D_s^-D^+$ decays. The measured value is
${\cal B}(\bbar\to D_s^-D^+)=\bigl[7.5\pm 0.2({\rm stat})\pm 0.8({\rm syst})\pm 0.8({\rm {\cal B}'s})\bigr]\times 10^{-3}$,  
which represents a large improvement in accuracy as compared to previous 
measurements~\cite{Bortoletto:1991kz, Albrecht:1991pa, Gibaut:1995tu, Aubert:2006nm}.
Combining this result with the world average for ${\cal B}(\bbar\to
D^-\pi^+)$~\cite{Yao:2006px} we obtain the ratio 
$R_{D_s/\pi}^{\rm ex.}/R_{D_s/\pi}^{\rm th.} = 0.85\pm 0.13(\rm ex.)\pm 0.09(\rm th.)$. With present experimental and
theoretical uncertainties, the results are consistent with the factorization hypothesis for non-leptonic exclusive decays of $B$ mesons.
If one does not include the penguin contributions~\cite{Kim:2001cj} to the amplitude for $\bdsd$ decay, the above ratio is not consistent with unity.
For $\bdsds$ decays we found no statistically significant signal. We set an
upper limit of
${\cal B}(\bdsds)\le 3.6\times 10^{-5}$ at 90\% C.L.
This result puts even more stringent limits on
${\cal B}(\bdsds)$ than the recent measurement by the BaBar
collaboration~\cite{Aubert:2005jv}, severely challenges recent theoretical 
estimates in Refs.~\cite{Li:2003az,Eeg:2005au} 
and implies that the weak annihilation contributions in
decay modes with two charmed mesons are small, as suggested in Ref.~\cite{Chen:2005rp}.

We thank the KEKB group for excellent operation of the
accelerator, the KEK cryogenics group for efficient solenoid
operations, and the KEK computer group and
the NII for valuable computing and Super-SINET network
support.  We acknowledge support from MEXT and JSPS (Japan);
ARC and DEST (Australia); NSFC and KIP of CAS (China); 
DST (India); MOEHRD, KOSEF and KRF (Korea); 
KBN (Poland); MIST (Russia); ARRS (Slovenia); SNSF (Switzerland); 
NSC and MOE (Taiwan); and DOE (USA).

\end{document}